
%
%
%
%
%
%
%
%
\input harvmac
\font\tay=eusb10
\font\tentit=cmmib10

\font\seventit=cmmib7
\font\fivetit=cmmib5
\newfam\titfam
\textfont\titfam=\tentit
\scriptfont\titfam=\seventit
\scriptscriptfont\titfam=\fivetit
\def\tit{\fam\titfam\tentit}
\def\CA{{\cal A}}   
 \def\CF{{\cal F}} \def\CG{{\cal G}} \def\CH{{\cal H}}
  \def\CK{{\cal K}} 
   \def\CP{{\cal P}}
  \def\CS{{\cal S}} 
 \def\CV{{\cal V}}  
 \def\CZ{{\cal Z}}
\def\rvec{{\bf \vec r}}
\def\kvec{{\bf \vec k}}
\def\Tay{{\hbox{\tay T}}}
\def\RR{\relax{\rm I\kern-.18em R}}
\def\Rr{\relax{\ninerm I\kern-.22em \ninerm R}}
\def\setminusp{\hbox{$/ \kern -3pt {}_p$}}
\def\ssetminusp{\hbox{$\scriptstyle / \kern -2pt {}_p$}}
\def\figcap#1#2{\nfig#1{#2} \noindent {Fig.\ }\xfig#1{:\ }{\sl #2}}
\def\figinsert#1#2#3{\vbox to #1{\vfill \vbox {\hsize=#2
\baselineskip=10pt {#3} \par}}}
\Title{SPhT/92-156 hep-th/9212102}{
\vbox{
\vskip -2truecm
\centerline{Renormalization of Crumpled Manifolds}
}}
\centerline{Fran\c cois David\footnote{$^\dagger$}
{Member of CNRS},
Bertrand Duplantier{$^\dagger$}{} and Emmanuel Guitter}
\bigskip\centerline{Service de Physique Th\'eorique\footnote{$^\star$}{
Laboratoire de la Direction des Sciences de la Mati\`ere du Commissariat
\`a l'Energie Atomique}}
\centerline{C.E. Saclay}
\centerline{F-91191 Gif-sur-Yvette, France}
\vskip 1.truecm
\centerline{\bf Abstract}{
\ninerm
\textfont0=\ninerm
\font\ninemit=cmmi9
\font\sevenmit=cmmi7
\textfont1=\ninemit
\scriptfont0=\sevenrm
\scriptfont1=\sevenmit
\baselineskip=11pt
\bigskip
We consider a model of $D$-dimensional tethered manifold interacting
by excluded volume in \Rr${}^d$ with a single point. By use of
intrinsic distance geometry, we first provide a rigorous definition of the
analytic continuation of its perturbative expansion for arbitrary $D$,
\ $0\!<\!D\!<\!2$. We then construct explicitly a renormalization operation
{\ninebf R}, ensuring renormalizability to all orders.
This is the first example of mathematical construction and renormalization
for an interacting extended object with continuous internal dimension,
encompassing field theory. \par
}
\vskip .3in
\Date{PACS numbers: 05.20.-y, 11.10.Gh, 11.17.+y}
\hfuzz 1.pt
\vfill\eject
The Statistical Mechanics of random surfaces and membranes,
or more generally of extended objects, poses fundamental problems
\ref\Jerus{{\sl Statistical Mechanics of Membranes and
Surfaces}, Proceedings of the Fifth Jerusalem Winter School for Theoretical
Physics (1987), D. R. Nelson, T. Piran and S. Weinberg Eds., World Scientific,
Singapore (1989).}.
\nref\SirSam{S. F. Edwards, Proc. Phys. Soc. Lond. {\bf 85} (1965) 613.}
\nref\desClJan{J. des Cloizeaux and G. Jannink,
{\sl Polymers in Solution, their Modelling and Structure}, Clarendon
Press, Oxford (1990).}
\nref\Wenetal{X. Wen {\it et al.},
Nature {\bf 355} (1992) 426.}
\nref\Spect{C.F. Schmidt {\it et al.},
unpublished.}
\hskip -20pt Among those, the study of {\it polymerized}
membranes, which are simple generalizations of linear polymers
[\xref\SirSam ,\xref\desClJan ] to two-dimensionally connected networks,
is prominent,
with a number of possible experimental
realizations [\xref\Wenetal ,\xref\Spect ]. From a theoretical point
of view, a clear challenge is to understand
self-avoidance (SA) effects in membranes.
\nref\KN{M. Kardar and D. R. Nelson, Phys. Rev. Lett. {\bf 58}
(1987) 1289, 2280(E); Phys. Rev. {\bf A 38}
(1988) 966.}
\nref\ArLub{J. A. Aronowitz and T. C. Lubensky, Europhys. Lett. {\bf 4}
(1987) 395.}
\kern -10pt Recently, a model was proposed [\xref\KN ,\xref\ArLub ]
which aimed to incorporate the advances made in polymer theory
by Renormalization Group methods into the field of polymerized, or
tethered membranes. These extended objects, a priori two-dimensional in nature,
are generalized for theoretical purposes to intrinsically
{\it $D$- dimensional manifolds} with internal points $x\in \RR^D$,
embedded in external $d$-dimensional space with position
vector $\rvec (x)\in \RR^d$.
The associated continuum Hamiltonian $\CH$ generalizes that of
Edwards for polymers \SirSam :
\eqn\Edwards{
\beta{\cal H}={1\over 2}\int d^Dx\,\Big(\nabla_x\rvec (x)\Big)^2+
{b\over 2}\int d^Dx\int d^Dx'\ \delta^{d}\big(\rvec (x)-\rvec (x')\big)
\ ,}
\nref\BDbis{B. Duplantier, Phys. Rev. Lett. {\bf 58} (1987) 2733;
and in \Jerus .}
\hskip -3pt with an elastic Gaussian term and a self-avoidance
two-body $\delta$-potential with interaction parameter $b>0$. For
$0<D<2$, the Gaussian manifold ($b=0$) is {\it crumpled} with a finite
Hausdorff dimension
$d_H=2D/(2-D)$; and the finiteness of the upper critical dimension
$d^\star = 2d_H$ for the SA-interaction
allows for an $\epsilon$-expansion about
$d^\star$ [\xref\KN --\nobreak\xref\BDbis ],
performed via a direct renormalization method
adapted from that of des Cloizeaux in polymer theory
\ref\desCloiz{J. des Cloizeaux, J. Phys. France {\bf 42} (1981) 635.}.

It should be stressed however that only the polymer case,
with an {\it integer} internal dimension $D=1$, can be
mapped,
following de Gennes \ref\DeGe{P.G. de Gennes, Phys. Lett.
{\bf A 38} (1972) 339.},
onto a standard field theory, namely a $({\bf \Phi}^2)^2$ theory
for a field ${\bf \Phi}$ with $n\to 0$ components. This
is instrumental to show that the direct renormalization
method for polymers is mathematically sound
\ref\BenMa{M. Benhamou and G. Mahoux, J. Phys. France {\bf 47} (1986) 559.},
and equivalent to rigorous renormalization schemes in standard
local field theory, such as the landmark
Bogoliubov Parasiuk Hepp Zimmermann (BPHZ) construction
\ref\BPHZ{N. N. Bogoliubov and O. S. Parasiuk, Acta Math. {\bf 97} (1957) 227;
\hfill\break\noindent
K. Hepp, Commun. Math. Phys. {\bf 2} (1966) 301;
\hfill\break\noindent
W. Zimmermann, Commun. Math. Phys. {\bf 15} (1969) 208.  }.
For manifold theory, we have to deal with {\it non-integer} internal
dimension $D$, $D\ne 1$, where no such mapping exists.
Therefore, two outstanding problems remain in the theory
of interacting manifolds: (a) the mathematical meaning of a
{\it continuous} internal dimension $D$;
(b) the actual {\it renormalizability} of the perturbative
expansion of a manifold model like \Edwards ,
implying scaling as expected on physical grounds.

A first answer was brought up in
\ref\BD{B. Duplantier, Phys. Rev. Lett. {\bf 62} (1989) 2337.},
where a simpler model
of a crumpled manifold interacting by excluded volume with
a fixed Euclidean subspace of $\RR^d$ was proposed. The direct
resummation of leading divergences of the perturbation series
indeed validates there {\it one-loop} renormalization,
a result later extended to the Edwards model \Edwards\
\ref\DHK{B. Duplantier, T. Hwa and M. Kardar, Phys. Rev. Lett. {\bf 64}
(1990) 2022.}.

In this Letter, we announce the results of an extensive study
of these questions
\ref\DDG{F. David, B. Duplantier and E. Guitter, Saclay preprint SPhT/92-124.}.
We first propose a mathematical
construction of the $D$-dimensional internal measure $d^Dx$
via distance geometry within the elastic manifold,
with expressions for manifold Feynman
integrals which generalize the $\alpha$-parameter representation
of field theory. In the case
of the manifold model of \BD\ , we then describe the essential
properties which make it indeed {\it renormalizable to all orders}
by a renormalization of the coupling constant,
and we directly construct a renormalization operation,
generalizing the BPHZ construction to manifolds.

The simplified model Hamiltonian introduced in \BD\ reads:
\eqn\HamM{
\beta{\cal H}={1\over 2}\int d^Dx\,\Big(\nabla_x \rvec (x)
\Big)^2+ {b}\int d^Dx\ \delta^{d}\big(\rvec (x)\big)
\ ,}
with now a pointwise interaction of the
Gaussian manifold with the origin.
Notice that this Hamiltonian also represents
interactions of a fluctuating (possibly directed) manifold
with a nonfluctuating $D'$- Euclidean subspace of $\RR^{d+D'}$,
$\rvec $ then standing for the coordinates transverse to this subspace.
The excluded volume case ($b>0$) parallels
that of the Edwards model \Edwards\ for SA-manifolds,
while an attractive interaction ($b<0$) is also possible, describing
pinning phenomena.
The dimensions of $\rvec$ and $b$ are respectively
$[\rvec ]=[x^\nu]$ with a size exponent $\nu\equiv (2-D)/2$, and
$[b]=[x^{-\epsilon}]$ with $\epsilon\equiv D-\nu d$.
For fixed $D$ and $\nu$, the parameter $d$ (or equivalently $\epsilon$)
controls
the relevance of the interaction, with the exclusion of a point
only effective for $d\le d^\star=D/\nu$.

The model is described by its (connected) partition function
$\CZ=\CV^{-1}\int {\cal D}[\rvec]\exp(-\beta{\cal H})$ (here $\CV$
is the internal volume of the manifold)
and, for instance, by its one-point vertex function
$\CZ^{(0)}(\kvec )/\CZ=
\int d^Dx_0 \langle e^{i\kvec\cdot\rvec(x_0)} \rangle $, where the
(connected) average
$\langle \cdots \rangle$ is performed with \HamM .
Those functions are formally defined via their perturbative expansions
in the coupling constant $b$:
$\CZ=\sum_{N=1}^{\infty}{(-b)^N\over N! }\,\CZ_N$ and a similar equation for
$\CZ^{(0)}$ with coefficients $\CZ_N^{(0)}$.
The term of order $N$, $\CZ_N$, is a ($b=0$) Gaussian average
involving $N$ interaction points $x_i$.
This average is expressed
solely in terms of the Green function
$G(x,y)=-{1\over 2}A_D|x\nobreak -\nobreak y|^{2\nu}$,
solution of $-\Delta_x G(x,y)=\delta^D (x-y)$,
with $A_D$ a suitable normalization, hereafter omitted.
In the following, it is important to preserve the condition
$0<\nu <1$ ({\it i.e.} $0<D<2$), corresponding to
the actual case of a  crumpled manifold, and where $(-G)$ is
positive and ultraviolet (UV) finite.
A direct evaluation of $\CZ_N$ then leads to its integral
representation in terms of the normalized
$G_{ij}\equiv - {1\over 2} |x_i-x_j|^{2\nu}$ \BD :
\eqn\ZN{
\CZ_N\ =\,{1\over \CV}\, \int\prod_{i=1}^N d^Dx_i
\,\left(\det\left[ \Pi_{ij}\right]_{\scriptscriptstyle 1\le i,j\le  N-1 }
\right)^{-{d\over 2}}
}
where the matrix $[\Pi_{ij}]$ is simply defined as
$\Pi_{ij}\,\equiv\,G_{ij}-G_{Nj}-G_{iN}$ with a
reference point, $x_N$, the symmetry between the $N$ points being
restored in the determinant.
The integral representation of $\CZ_N^{(0)}$ is obtained from
that of $\CZ_N$ by multiplying the integrand in \ZN\ by
$\exp (-{1\over 2}\kvec^2 \Delta^{(0)} )$ with :
\eqn\forvertex{\Delta^{(0)}\equiv {\det[\Pi_{ij}]_{\scriptscriptstyle
0\le i,j \le N-1} \over
\det[\Pi_{ij}]_{\scriptscriptstyle 1\le i,j\le N-1} } \ ,}
and integrating over one more position, $x_0$.
The resulting expression is quite similar to that
of the manifold Edwards model \DHK .

\medskip
{\bf Analytic continuation in ${\tit D}$ of the Euclidean measure}.
Integrals like \ZN\ are {\it a priori} meaningful only for
integer $D$. Still, an analytic continuation in $D$ can be performed
by use of {\it distance geometry}. The key idea is to substitute
to the internal Euclidean coordinates $x_i$ the set of all mutual (squared)
distances $a_{ij}=(x_i-x_j)^2$. This is possible for integrands invariant
under the group of Euclidean motions (as in \ZN\ and \forvertex ). For
$N$ integration points, it also requires $D$ large enough, {\it i.e.}
$D\ge N-1$, such that $N-1$ relative vectors spanning these points
are linearly independent.
We define the graph $\CG$ as the set $\CG =\{1,\ldots ,N\}$
labelling the interaction points.
Vertices $i\in\CG$ will be remnants of the original Euclidean points
after analytic continuation, and index the distance matrix $[a_{ij}]$.
The change of variables $\{x_i\}_{i\in \CG}
\to a\equiv [a_{ij}]_{{i<j \hfill \atop i,j \in \CG}}$
reads explicitly \DDG :
\eqn\inta{
{1\over \CV}\int_{\RR^D} \prod_{i\in \CG} d^Dx_i\,\cdots \ =\
\int_{{\cal A}_\CG}\, d\mu_\CG^{(D)}(a)\,\cdots
\ ,}
with the measure
\eqn\measaij{
d\mu_\CG^{(D)}(a)\equiv
\prod_{{i<j  \atop i,j \in \CG}}da_{ij}\
\Omega_N^{(D)}\, \Big(P_\CG(a) \Big)^{D-N\over 2}
\ ,}
where $N=|\CG|$,
$\Omega_N^{(D)}\equiv \prod_{K=0}^{N-2} {S_{D-K}\over 2^{K+1}}$
(here $S_D={2\pi^{D/2}\over \Gamma(D/2)}$
is the volume of the unit sphere in $\RR^D$), and
\eqn\CayMen{
P_\CG(a)\equiv {(-1)^N \over 2^{N-1}}\,
\left| \matrix{0&1&1&\ldots&1\cr 1&0&a_{12}&\ldots&a_{1N}\cr
1&a_{12}&0&\ldots&a_{2N}\cr \vdots&\vdots&\vdots&\ddots&\vdots\cr 1&a_{1N}&
a_{2N}&\ldots&0\cr } \right|\ . }
The factor
$\Omega_N^{(D)}$ is the volume of the rotation group of the rigid
simplex spanning the points $x_i$. The ``Cayley-Menger determinant"
\ref\Blum{L. M. Blumenthal, {\sl Theory and Applications of Distance Geometry},
Clarendon Press, Oxford (1953).}
$P_\CG(a)$ is proportional to the squared Euclidean
volume of this simplex, a polynomial of degree $N-1$ in the $a_{ij}$.
The set $a$ of squared distances has to fulfill the triangular
inequalities and their generalizations:
$P_\CK(a)\ge 0$ for all subgraphs $\CK\subset \CG$,
which defines the domain of integration $\CA_\CG$ in \inta .
For real $D>|\CG|-2$, $d\mu_\CG^{(D)}(a)$ is a positive measure on $\CA_\CG$,
analytic in $D$.
It is remarkable that, as a distribution, it can be extended
to $0\le D\le |\CG|-2$ \DDG . For integer $D\le |\CG|-2$,
although the change of
variables from $x_i$ to $a_{ij}$ no longer exists, Eq.\measaij\
still reconstructs the correct measure, concentrated on
$D$-dimensional submanifolds of $\RR^{N-1}$,
{\it i.e.} $P_\CK=0$ if $D\le |\CK|-2$
\DDG .
For example, when $D\to 1$ for $N=3$ vertices, we have,
denoting the distances $|ij|=\sqrt{a_{ij}}$:
$${d\mu^{(D\to 1)}_{\{1,2,3\}}(a)\over
d{\scriptstyle |12|}
d{\scriptstyle |13|}
d{\scriptstyle |23|}}
=2\, \delta\big({\scriptstyle |12|+|23|-|13|}\big)
+\hbox{\ninerm perm}
$$
which indeed describes nested intervals in $\RR$.
\medskip
Another nice feature of this formalism is that
the interaction determinants in \ZN\ and \forvertex\ are
themselves Cayley-Menger determinants. We have indeed
$\det\left[\Pi_{ij}\right]_{1\le i,j \le N-1} = P_\CG(a^\nu)$
where $a^\nu\equiv[a_{ij}^\nu]_{{i<j\hfill \atop i,j \in \CG}}$ is obtained
by simply raising
each squared distance to the power $\nu$.
We arrive at the representation of ``Feynman diagrams" in
distance geometry:
\eqn\ZNa{\eqalign{
&\CZ_N=\int_{\CA_\CG}d\mu^{(D)}_\CG\, I_\CG\ ,\ \ \
I_{\CG} =\big(P_\CG(a^\nu)\big)^{ -{d\over 2}}  \cr
&\CZ_N^{(0)}=
\int_{\CA_{\CG \cup \{0\}}}d\mu^{(D)}_{\CG \cup \{0\}}
\, I_\CG^{(0)} \ ,\cr
&I_{\CG}^{(0)} =
I_\CG \ \exp \left(-{1\over 2}\kvec^2
{P_{\CG\cup\{0\}}(a^\nu)\over P_\CG(a^\nu)}
\right)\hfill ,\cr } }
which are $D$-dimensional extensions of the Schwinger $\alpha$-parameter
representation.
We now have to study the actual
convergence of these integrals and, possibly, their renormalization.
\medskip
{\bf Analysis of divergences}.
Large distance infrared (IR) divergences occur for
manifolds of infinite size. One can keep a
finite size, preserve symmetries and
avoid boundary effects by choosing as a manifold
the $D$-dimensional sphere $\CS_D$ of radius $R$
in $\RR^{D+1}$. This amounts \DDG\ in distance geometry
to substituting to $P_\CG(a)$ the ``spherical" polynomial
$P_\CG^{\CS}(a)\equiv  P_\CG(a)+{1\over R^2}\det(-{1\over 2}a)$,
the second term providing an IR cut-off, such that $a_{ij}\le 4R^2$.
In the following, this regularization
will be simply ignored when dealing with
short distance properties, where $P_\CG^\CS\sim P_\CG$.
\medskip
{\it Schoenberg's theorem}. This result of geometry \Blum\ states that
{\sl for $0<\nu<1$, the set
$a^\nu=[a^\nu_{ij}]_{{ i<j\hfill \atop i,j \in \CG}}$ can be realized as the
set of
squared distances
of a transformed simplex in $\RR^{N-1}$, whose volume
$P_\CG(a^\nu)$ is positive and vanishes if and only if at least one of
the mutual
original distances itself vanishes}, $a_{ij}=0$.
This ensures that, as in field theory, the only source
of divergences in $I_\CG$ and $I_\CG^{(0)}$ is at {\it short distances}.
Whether these UV singularities are integrable or not will depend
on whether the external space dimension $d< d^\star
=D/\nu$ or $d>d^\star$.

{\bf Factorizations}. The key to convergence and
renormalization is the following
short distance {\it factorization} property of $P_\CG(a^\nu)$.
Let us consider a subgraph $\CP\subset\CG$, with at least two vertices,
in which we
distinguish an element, the {\it root} $p$ of $\CP$, and let us denote by
$\CG\setminusp\CP \equiv (\CG\setminus\CP )\cup \{p\}$ the subgraph
obtained by replacing in $\CG$ the whole
subset $\CP$ by its root $p$.
In the original Euclidean formulation, the analysis of short distance
properties amounts to that of contractions of points $x_i$, labeled
by such a subset $\CP$, toward the point $x_p$, according to:
$x_i(\rho )=x_p+\rho (x_i-x_p)$ if $i\in \CP$,
where $\rho\to 0^{+}$ is the dilation factor, and
$x_i(\rho)=x_i$ if $i\notin \CP$.
This transformation has
an immediate correspondent
in terms of mutual distances:
$a_{ij}\ \to a_{ij}(\rho)$, depending on both $\CP$ and $p$.
Under this transformation, the interaction polynomial
$P_\CG(a^\nu)$ factorizes into \DDG :
\eqn\factdet{\eqalign{
P_\CG(a^\nu (\rho)) & =
P_\CP(a^\nu(\rho ))\,
P_{\CG\setminusp \CP}(a^\nu) \cr & \quad \quad  \times
\left\{1+{\cal O}(\rho^{2\delta})\right\}\
\ .\cr}}
with $ \delta=\min(\nu, 1-\nu)>0 $ and
where, by homogeneity,
$P_\CP(a^\nu(\rho ))=\rho^{2\nu (|\CP |-1)}\, P_\CP(a^\nu)$.
\midinsert
\figinsert{6.5truecm}{7.truecm}{
\figcap\contf{Factorization property \factdet . }}
\includegraphics{fig.ps}
\endinsert
\noindent The geometrical interpretation of \factdet\ is quite
simple:
the contribution of the set $\CG$ splits into that of
the contracting subgraph $\CP$ multiplied by that of the whole set $\CG$ where
$\CP$ has been replaced by its root $p$ (\contf ),
all correlation distances between these subsets being suppressed.
This is just, in this model,
the rigorous expression of an {\it operator
product expansion} \DDG .

The factorization property \factdet\ does not hold for $\nu = 1$,
preventing a factorization of the measure \measaij\
$d\mu^{(D)}_{\CG }(a)$ itself.
Still, the integral of the measure, when applied to a factorized
integrand, factorizes as:
\eqn\factint{\int_{\CA_\CG}d\mu^{(D)}_\CG\cdots=\int_{\CA_\CP}
d\mu^{(D)}_\CP\cdots\int_{\CA_{(\CG\ssetminusp \CP)}}
d\mu^{(D)}_{(\CG\setminusp\CP)}\cdots \ .}
This fact, explicit for integer $D$ with a
readily factorized measure $\prod_i d^Dx_i$,
is preserved \DDG\ by analytic
continuation only after integration over relative distances between the
two ``complementary" subsets $\CP$ and $\CG\setminusp \CP$.
\medskip
{\bf Renormalization}.
A first consequence of factorizations \factdet\ and \factint\ is
the absolute convergence of $\CZ_N$ and $\CZ_N^{(0)}$ for $\epsilon >0$.
Indeed, the superficial degree of divergence of $\CZ_N$ (in distance units)
is $(N-1)\epsilon$, as can be read from \ZNa , already ensuring the
superficial convergence when $\epsilon>0$.
The above factorizations ensure that the superficial degree of
divergence in $\CZ_N$ or $\CZ_N^{(0)}$ of any subgraph $\CP$ of $\CG$
is exactly that of $\CZ_{|\CP |} $ itself, {\it i.e.} $(|\CP |-1)\epsilon >0$.
By recursion, this ensures the absolute convergence of the manifold
Feynman integrals. A complete discussion has recourse to a generalized
notion of Hepp sectors and is given elsewhere \DDG . In the proof,
it is convenient to first consider $D$ large enough
where $d\mu_{\CG}^{(D)}$ is a non singular measure,
with a fixed $\nu$ considered as an independent variable $0<\nu<1$,
and to then continue to $D=2-2\nu$, $0<D<2$,
corresponding to the physical case.

When $\epsilon =0$, the integrals giving $\CZ_N$ and $\CZ_N^{(0)}$
are (logarithmically) divergent.
Another main consequence of Eqs. \factdet\ and \factint\ is then
the possibility
to devise a renormalization operation {\bf R},
as follows.
To each contracting rooted subgraph $(\CP,p)$ of $\CG$,
we associate a Taylor operator $\Tay_{(\CP,p)}$, performing
on interaction integrands the exact factorization corresponding to \factdet :
\eqn\TayI{\Tay_{(\CP,p)}I_\CG^{(0)}=I_{\CP}\,I_{\CG\setminusp \CP}^{(0)}
\ ,}
and similarly
$\Tay_{(\CP,p)}I_\CG=I_{\CP}\,I_{\CG\setminusp \CP}$.
As in standard field theory \BPHZ ,
the subtraction renormalization operator {\bf R}
is then organized in terms of forests \`a la Zimmermann.
In manifold theory, we define
a {\it rooted forest} as a set of rooted subgraphs $(\CP,p)$ such
that any two subgraphs are either disjoint or nested, {\it i.e.} never
partially overlap. Each of these subgraphs in the forest
will be contracted toward its root under the action \TayI\ of the
corresponding Taylor operator.
When two subgraphs $\CP\subset \CP'$ are nested, the smallest one
is contracted first toward its root $p$, the root
$p'$ of $\CP'$ being itself attracted toward $p$
if $p'$ happened to be in $\CP$.
This hierarchical structure is anticipated by choosing the roots of the
forest as {\it compatible}: in the case described above, if $p'\in \CP$,
then $p'\equiv p$.
Finally, the renormalization operator is written
as a sum over all such compatibly rooted forests of $\CG$, denoted by
$\CF_\oplus$:
\eqn\Roper{
{\bf R}\ =\
\sum_{\CF_{\oplus}}\,W(\CF_{\oplus})\,
\Bigg[ \prod_{(\CP,p)\in \CF_{\oplus}}
\!\big( -\Tay_{(\CP,p)} \big)\Bigg]
\ .}
Here $W$ is a necessary combinatorial weight associated with the degeneracy
of compatible rootings, $W(\CF_{\oplus})\,
=\, \prod_{ {p\ {\rm root}}\atop {{\rm of}\,\CF_{\oplus}} }
1 /|\CP(p)| $ with $\CP(p)$ being the largest subgraph of the forest
$\CF_\oplus$ whose root is $p$.
An important property is that, with compatible roots, the Taylor operators
of a given forest now commute \DDG .
The renormalized amplitudes are defined as
\eqn\ZMRen{
{\CZ^{\bf R}}_N^{(0)}\ \equiv \
\int_{\CA_{\CG \cup \{0\}}} d\mu^{(D)}_{\CG\cup\{0\}}\,{\bf R}\,[I_\CG^{(0)}]
\ .}
The same operation ${\bf R}$
acting on $I_\CG$ leads automatically by homogeneity to
$ {\bf R}\left[I_{\CG}\right]=0 $ for $|\CG|\ge 2$.
We state the essential result that now {\it the renormalized Feynman
integral} \ZMRen\ {\it is convergent}: ${\CZ^{{\bf R}}}_N^{(0)} < \infty$
for $\epsilon=0$.
A complete proof of this renormalizability property goes
well beyond the scope of this Letter and is given elsewhere \DDG .
the analysis being
inspired from the direct proof by Berg\`ere and Lam of the
renormalizability in field theory of Feynman amplitudes
in the $\alpha$-representation
\ref\BergLam{M. C. Berg\`ere and Y.-M. P. Lam,
J. Math. Phys. {\bf 17} (1976) 1546.}.
\medskip
The physical interpretation of the renormalized amplitude \ZMRen\ and
of \Roper\ is now fairly simple.
Eqs.\factint\ and \TayI\ show that
the substitution to the bare amplitudes \ZNa\ of the renormalized ones
\ZMRen\ amounts to a reorganization to all orders of the original
perturbation series in $b$,
leading to the remarkable identity:
\eqn\RenExp{
\CZ^{(0)}\ =\ \sum_{N=1}^\infty \,{(\CZ)^N\over N!}\,
{\CZ^{\bf R}}_N^{(0)}
\ .}
This actually extends
to any vertex function, showing that the theory is
made perturbatively finite (at $\epsilon = 0$)
by a simple renormalization of the coupling constant $b$
into $\CZ$ itself.  From this result,
one establishes the existence of a Wilson function
${\scriptstyle \CV} {\partial \CZ \over \partial \CV} \big|_b$,
describing the scaling properties of the interacting manifold
for $\epsilon$ close to zero \DDG .
For $\epsilon >0$, an IR
fixed point at $b>0$ yields universal excluded volume exponents;
for $\epsilon <0$, the associated UV fixed point at $b<0$ describes
a localization transition.

In summary, we have shown how to define an interacting manifold model with
continuous internal dimension, by use of distance geometry, a natural
extension of Schwinger representation of field theories. Furthermore,
in the case of a pointwise interaction, we have shown that the
manifold model is indeed renormalizable to all orders.
The main ingredients are the Schoenberg's theorem
of distance geometry, insuring that divergences occur only at short
distances for (finite) manifolds, and the short-distance factorization
of the generalized Feynman amplitudes. The renormalization operator
is a combination
of Taylor operators associated with rooted diagrams,
a specific feature of manifold models.
This is probably the
first example of a perturbative renormalization
established for extended geometrical objects.
This opens the way to a similar study of self-avoiding manifolds,
as well as to other generalizations of field theories.
\medskip
We thank M. Berg\`ere for helpful discussions.

\vfill\eject
\listrefs
\bye